\input{aipcheck}

\documentclass[
    ,final            % use final for the camera ready runs
  ]
  {aipproc}

\layoutstyle{6x9}

\begin{document}

\title{Three-Body Scattering without Partial Waves}

\author{H. Liu}{
  address={Institute of Nuclear and Particle Physics, Ohio University, Athens, OH 45701, USA}
}

\author{Ch. Elster}{
  address={Institute of Nuclear and Particle Physics, Ohio University, Athens, OH 45701, USA}
}

\author{W. Gl\"ockle}{
  address={Institute for Theoretical Physics II, Ruhr-University Bochum, D-44780 Bochum, Germany}
}

\begin{abstract}
The Faddeev equation for three-body scattering at arbitrary energies is
formulated in momentum space and directly solved in terms of momentum vectors
without employing a partial wave
decomposition. In its simplest form the Faddeev equation for
identical bosons is a three-dimensional
integral equation in five variables, magnitudes of relative momenta and angles.
The elastic differential cross section, semi-exclusive d(N,N') cross sections and
total cross sections of both
elastic and breakup processes in the intermediate energy range up to about 1 GeV
are calculated based on a Malfliet-Tjon type potential,
 and the convergence of the multiple scattering series is investigated
in every case. In general a truncation in the first or second order in the two-body t-matrix is
quite insufficient.
\end{abstract}

\maketitle

%%%%%%%%%%%%%%%%%%%%%%%%%%%%%%%%%%%%%%%%%%%%
%% MAINMATTER
%%%%%%%%%%%%%%%%%%%%%%%%%%%%%%%%%%%%%%%%%%%%
%
%\input{psfig}

Traditionally three-nucleon scattering calculations are carried out by
solving Faddeev equations in a partial wave truncated basis. A partial
wave decomposition replaces the continuous angle variables by discrete
orbital angular momentum quantum numbers, and thus reduces the number
of continuous variables, which have to be discretized in a numerical
treatment. For low projectile energies the procedure of considering
orbital angular momentum components appears physically justified due
to arguments related to the centrifugal barrier. 
If one considers three-nucleon scattering at a few hundred
MeV projectile energy, the number of partial waves needed to achieve
convergence proliferates, and limitations with respect to
computational feasibility and accuracy are reached. The amplitudes
acquire stronger angular dependence, which is already visible in the
two-nucleon amplitudes, and their formation by an increasing number of
partial waves not only becomes more tedious but also less
informative. The method of partial wave decomposition looses its
physical transparency, and the direct use of angular variables becomes
more appealing.
It appears therefore natural to avoid a partial wave representation
completely and work directly with vector variables. 

Here we want to show that the full solution of the three-body scattering equation can be
obtained in a straightforward manner, when employing vector variables, i.e. magnitudes of momenta
and angles between the momentum vectors. As a simplification we neglect spin and isospin degrees
of freedom and treat three-boson scattering. The interaction employed is of Malfliet-Tjon type,
i.e. consists of a short range repulsive and intermediate range attractive Yukawa force.  
The parameters of the potential are adjusted so that a two-body bound state at $E_d$ = -2.23~MeV is
supported. Technical details of the calculation are given in \cite{scatter3d,hangthesis}. 

As first result we present in Fig.~1 the angular distribution for two energies, 0.2 and 1.0~GeV.
In the upper panels the entire angle range is displayed using a logarithmic scale, whereas the
lower panels focus on the forward angles. 

\leftskip=-0.1cm

\begin{minipage}[b]{8.0cm}
  \includegraphics[width=70mm,angle=-90]{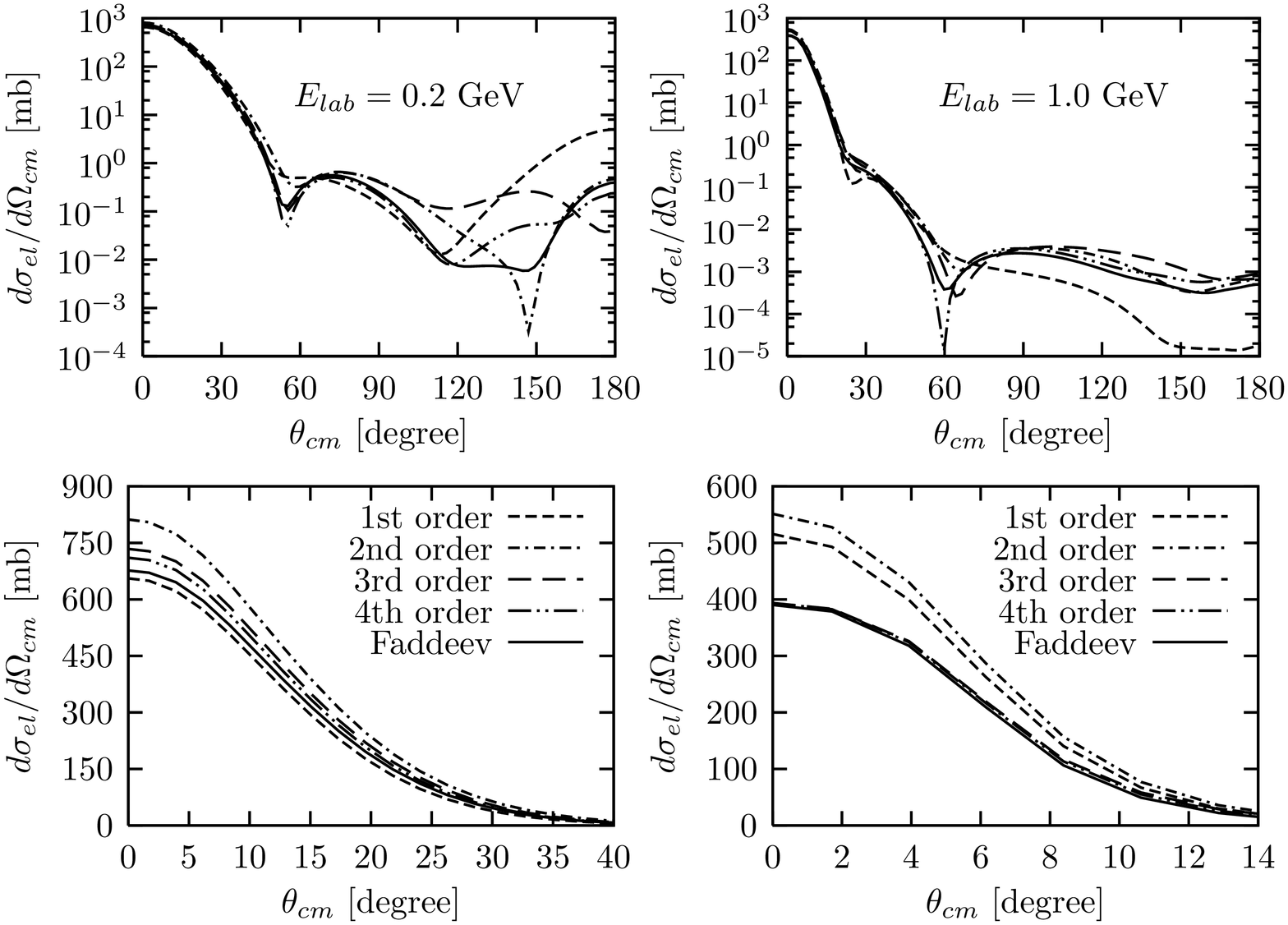}
\end{minipage}

\vspace*{-6.5cm}\leftskip=9.5cm
\begin{minipage}[h]{4.0cm}
{{\bf Fig.1: }\footnotesize
The elastic differential cross section at 0.2 and 1.0~GeV projectile energy as
function of the laboratory scattering angle. The solid line represents the full solution of the
Faddeev equation, whereas the other lines represent different orders in the multiple scattering
series as indicated in the legends of the lower panels.
}
\end{minipage}

\leftskip=0cm \vspace*{0.5cm}

\medskip \smallskip

In addition to the exact Faddeev result, the cross
sections are evaluated in first order in the two-body t-matrix, second order in $t$, third and
4th order in $t$ and successively added up as indicated. This allows to study the convergence of
the multiple scattering series. As expected, for the low energy, 0.2~GeV, rescattering terms of
higher order are important, and even the 4th order is not yet close to the exact result. This is
especially drastic for the large angles. At 1~GeV two rescattering terms (3rd order in $t$) are
necessary to come into the vicinity of the final result in forward direction. For the large
angles, the first rescattering correction has the biggest effect.

\leftskip=-0.1cm
\begin{minipage}[t]{8.0cm}
 \includegraphics[width=80mm,angle=-90]{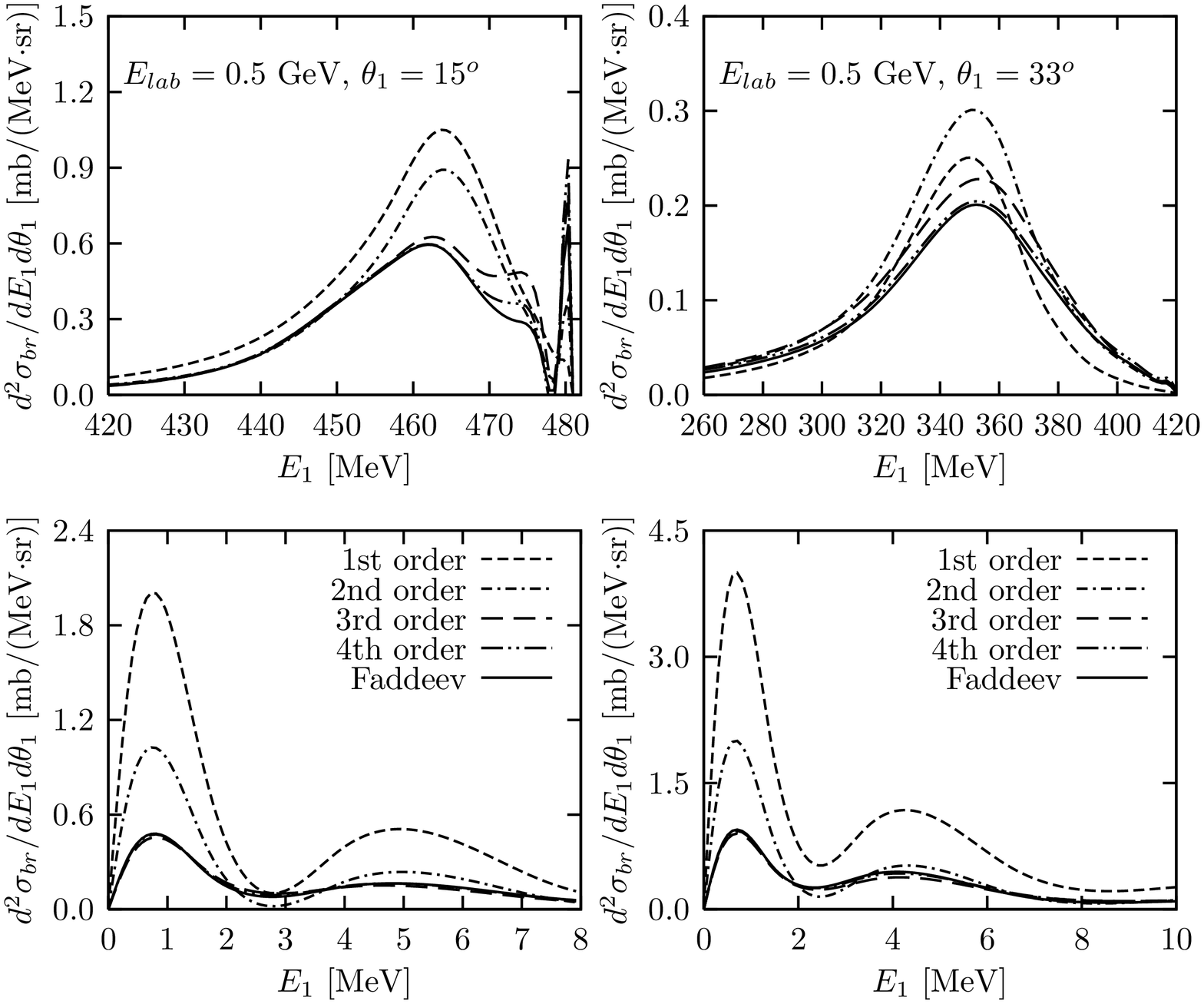}
\end{minipage}

\vspace*{-7.8cm}\leftskip=9.2cm
\begin{minipage}[h]{4.8cm}
{{\bf Fig.2: }\footnotesize
The semi-exclusive cross section at 0.5~GeV laboratory incident energy and at 15$^o$ angle
(left panels) and 33$^o$ angle of the emitted particle (right panels). In both cases the upper
panel displays the high energy range of the emitted particle, whereas the lower panel shows the
low energy range. The full solution of the Faddeev equation is given by the solid line in all
panels, The contribution of the lowest orders of the
multiple scattering series added up successively is given by the other curves as indicated in
the legends.
}
\end{minipage}

\leftskip=0cm \vspace*{0.2cm}

%\medskip

The semi-exclusive cross section d(N,N') for scattering at 0.5~GeV is given in Fig.~2 for the emission
angles 15$^o$ and 33$^o$. The upper panels show the high energies and the lower panels the low
energies of the emitted particle. Together with the full solution of the Faddeev equation (solid line)
the sums of the lowest orders of the multiple scattering series are shown as indicated in the
figure. The peak at the highest energy of the emitted particle is the so called final state
interaction (FSI) peak, which only develops if rescattering terms are taken into account. This
peak is a general feature of semi-exclusive scattering and is present at all energies. The next
peak is the so called quasi-free (QFS) peak, and one observes that at both angles one needs at least
rescattering up to the 3rd order to come close to the full result. At both angles the very low
energies of the emitted particle exhibit a strong peak in first order, which is considerably
lowered by the first rescattering. Here the calculation up to 3rd order in the multiple scattering
series seems already sufficient.

\leftskip=-0.1cm

\begin{minipage}[h]{8.0cm}
 \includegraphics[width=80mm,angle=-90]{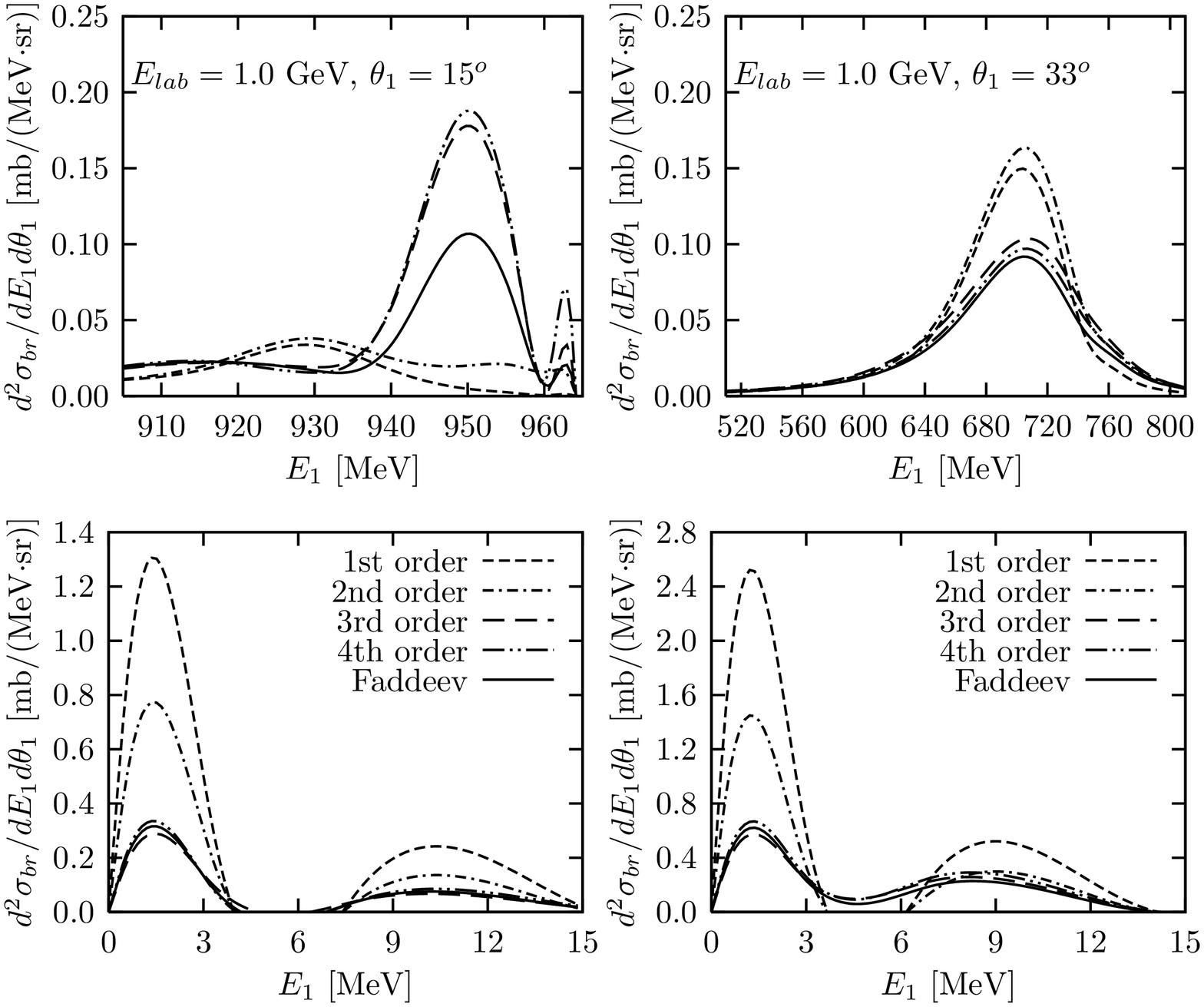}
\end{minipage}

\vspace*{-7.8cm}\leftskip=9.2cm
\begin{minipage}[h]{4.8cm}
{{\bf Fig.3: }\footnotesize
The semi-exclusive cross section at 1.0~GeV laboratory incident energy and at 15$^o$ angle
(left panels) and 33$^o$ angle of the emitted particle (right panels). In both cases the upper
panel displays the high energy range of the emitted particle, whereas the lower panel shows the
low energy range. The full solution of the Faddeev equation is given by the solid line in all
panels, The contribution of the lowest orders of the
multiple scattering series added up successively is given by the other curves as indicated in
the legends.
}
\end{minipage}

\leftskip=0cm\vspace*{0.5cm}

%\medskip

At 1~GeV the situation is similar, as shown in Fig.~3. It is interesting to observe that the
strong QFS peak for the small angle (15$^o$) only develops after two rescatterings (3rd order),
and the final height of the peak requires orders higher than the ones shown. 
For the larger angle (33$^o$) the third order calculation is already quite close to the exact
solution. Again the final
result for the peak at the very low energy of the emitted particle is reached with two rescattering
contributions.

In conclusion we can say that the three-body Faddeev equation can be safely solved at intermediate
energies using momentum vectors directly. Our present calculations are based on local forces,
however this is not a restriction of our approach. To the best of our knowledge these are the
first calculations of this kind. The key point is to neglect the partial wave decomposition
generally used at low energies. Thus all partial waves are exactly included. 
%We find that in
%nearly all cases studied processes of at least 2nd and 3rd order rescattering are required.
%Further studies scanning the complete three-body phase space for the total break-up are under
%way. This may be important to shed light on previous theoretical analysis of p(d,ppn) reactions
%which relied on low order reaction mechanisms. 

\vspace{-8mm}

\begin{theacknowledgments}
\vspace{-3mm}
This work was performed in part under the
auspices of the U.~S.  Department of Energy under contract
No. DE-FG02-93ER40756l with Ohio University. We thank
the National Energy Research Supercomputer Center (NERSC) for the use of
their facilities.
\end{theacknowledgments}

\end{document}